\pdfoutput=1
\documentclass[conference]{IEEEtran}

\IEEEoverridecommandlockouts

\usepackage{graphicx}
\usepackage{color}

\usepackage{cite}
\usepackage{amsmath}
\usepackage[caption=false]{subfig}

\usepackage{amssymb}
\usepackage{amsfonts}

\usepackage{algorithmic}
\usepackage{algorithm}

\usepackage{bbm}

\usepackage{comment}

\newtheorem{proposition}{\bf Proposition}

\newcommand{\zone}{z}
\newcommand{\ZONE}{Z}
\newcommand{\zoneSet}{\set{\ZONE}}

\newcommand{\vue}{u}
\newcommand{\VUE}{U}
\newcommand{\vueSet}{\set{\VUE}}

\newcommand{\coordinatesTx}{(x_{\vue},y_{\vue})}
\newcommand{\coordinatesRx}{(x_{\vue'},y_{\vue'})}
\newcommand{\pathloss}{\text{PL}}
\newcommand{\pathlossCoefficient}{\ell}
\newcommand{\pathlossExponent}{c}
\newcommand{\distanceBound}{d_0}

\newcommand{\rb}{n}
\newcommand{\RB}{N}
\newcommand{\rbSet}{\set{\RB}}

\newcommand{\arrival}{a}
\newcommand{\arrivalAvg}{\overline{\arrival}}

\newcommand{\queue}{q}
\newcommand{\queueTH}{\queue_0}

\newcommand{\queueMAXvar}{m}
\newcommand{\queueAvg}[1]{\expect[\queue_{#1}]}
\newcommand{\outage}{\epsilon}

\newcommand{\rate}{r}
\newcommand{\channel}{h}
\newcommand{\channelVec}{\vect{\channel}}

\newcommand{\noiseAlone}{ N_0}
\newcommand{\noise}{\bandwidth \noiseAlone}
\newcommand{\bandwidth}{W}

\newcommand{\txpower}{p}
\newcommand{\txpowerVec}{\vect{\txpower}}
\newcommand{\txpowerMax}{\txpower_0}
\newcommand{\interference}{I}
\newcommand{\interferenceEST}{\tilde{\interference}}

\newcommand{\gevCombined}{\vect{d}}

\newcommand{\loglikelihood}{f^{\gevCombined}}
\newcommand{\loglikelihoodSP}[1]{f^{#1}}

\newcommand{\gpdScale}{\sigma}
\newcommand{\gpdShape}{\xi}
\newcommand{\queuePoT}{M}

\newcommand{\gpd}{G_{\queuePoT}}
\newcommand{\gpdExponent}[1]{g^{\gevCombined}(#1)}
\newcommand{\gpdCombined}{\vect{d}}
\newcommand{\gpdCombinedFeasible}{\set{D}}
\newcommand{\gpdML}{G^{\gevCombined}_{X}}

\newcommand{\timeblock}{T}

\newcommand{\si}{\alpha}
\newcommand{\cinr}{\gamma}
\newcommand{\dualPower}{\lambda}

\newcommand{\vqQ}{\Upsilon}
\newcommand{\vqEXq}{A}

\newcommand{\lyapunov}[1]{L(#1)}
\newcommand{\lyapunovDrift}{\Delta L_t}
\newcommand{\queueCombined}{\Xi}
\newcommand{\lyapunovBound}{\Delta_0}
\newcommand{\lyapunovConst}{\Delta_{\vue}}
\newcommand{\lyapunovTradeoff}{V}

\newcommand{\block}{k}
\newcommand{\BLOCK}{K}

\newcommand{\blocklength}{w}
\newcommand{\sample}{Q}
\newcommand{\sampleSet}{\set{\sample}}
\newcommand{\blockTimeSet}{\set{T}}
\newcommand{\sampleSizeRatio}{\kappa_{\vue}}

\newcommand{\federatedTime}{T_f}
\newcommand{\stepsize}{\delta}

\newcommand{\txr}{\text{vTx-vRx}}%

\newcommand*{\myfigfactor}{0.95}

\DeclareMathOperator*{\argmin}{arg\,min}

\newcommand{\expect}{\mathbb{E}\,}
\newcommand{\probability}{\text{Pr}}

\newcommand{\vect}{\boldsymbol}

\newcommand{\seta}[1]{1,\dots,#1}
\newcommand{\set}[1]{\mathcal{#1}}
\newcommand{\setSize}[1]{|#1|}

\newcommand{\vectx}{\vect{x}}

\newcommand{\one}{\mathbf{1}}
\newcommand{\zero}{\mathbf{0}}

\newcommand{\indictsimp}[1]{\mathbbm{1}_{#1}}

\newcommand{\transpose}{^\dag}
\newcommand{\optimal}{^\star}

\newcommand{\grad}[2]{\nabla_{#1}#2}
\newcommand{\daba}[2]{\frac{\partial #1}{\partial #2}}

\newcommand{\norm}[2]{\| #1 \|_{#2}}

\usepackage{forloop}
\newcounter{loopcntr}
\newcommand{\rpt}[2][1]{%
	\forloop{loopcntr}{0}{\value{loopcntr}<#1}{#2}%
}

\newcommand{\subgroup}[1]%
{\rlap{\smash{%
	\newcount\cnt%
	\cnt \numexpr#1\relax%
	\advance\cnt -1\relax%
	$\tabcolsep=.1em\begin{tabular}[t]{|l}\multicolumn{1}{l}{}\\%
	\rpt[\cnt]{\\}
	\\\hline\end{tabular}$%
}}}

\newcounter{myRefCount}

\includecomment{comment}

\begin{document}

\title{%
	Federated Learning for Ultra-Reliable Low-Latency V2V Communications 
}

\author{
\IEEEauthorblockN{
	Sumudu Samarakoon\IEEEauthorrefmark{1},
	Mehdi Bennis\IEEEauthorrefmark{1},
	Walid Saad\IEEEauthorrefmark{2},
	and M\'{e}rouane Debbah\IEEEauthorrefmark{3}
	\\}
\IEEEauthorblockA{
	\small%
	\IEEEauthorrefmark{1} Centre for Wireless Communication, University of Oulu, Finland, email: \{sumudu.samarakoon,mehdi.bennis\}@oulu.fi \\
	\IEEEauthorrefmark{2}Wireless@VT, Bradley Department of Electrical and Computer Engineering, Virginia Tech, Blacksburg, VA, email: walids@vt.edu \\
	\IEEEauthorrefmark{3} Mathematical and Algorithmic Sciences Lab, Huawei France R\&D, Paris, France, (email: merouane.debbah@huawei.com)
}
}

\maketitle
\nopagebreak[4]
\begin{abstract}

In this paper, a novel joint transmit power and resource allocation approach for enabling ultra-reliable low-latency communication (URLLC) in vehicular networks is proposed. 
The objective is to minimize the network-wide power consumption of vehicular users (VUEs) while ensuring high reliability in terms of probabilistic queuing delays. 
In particular, a reliability measure is defined to characterize extreme events (i.e., when vehicles' queue lengths exceed a predefined threshold with non-negligible probability) using extreme value theory (EVT). 
Leveraging principles from federated learning (FL), the distribution of these extreme events corresponding to the tail distribution of queues is estimated by VUEs in a decentralized manner.
Finally, Lyapunov optimization is used to find the joint transmit power and resource allocation policies for each VUE in a distributed manner. 
The proposed solution is validated via extensive simulations using a Manhattan mobility model.
It is shown that FL enables the proposed distributed method to estimate the tail distribution of queues with an accuracy that is very close to a centralized solution with up to 79\% reductions in the amount of data that need to be exchanged.
Furthermore, the proposed method yields up to 60\% reductions of VUEs with large queue lengths, without an additional power consumption, compared to an average queue-based baseline.
Compared to systems with fixed power consumption and focusing on queue stability while minimizing average power consumption, the reduction in extreme events of the proposed method is about two orders of magnitude.

\end{abstract}

\section{Introduction}\label{sec:introduction}

Vehicle-to-vehicle (V2V) communication is a key enabler for autonomous driving and intelligent transportation systems \cite{pap:ikram17,jnl:liu18,jnl:zeng18,tech:5gcar,jnl:shah18}.
\begin{comment}
The ability to extend vehicles’ field-of-view improves traffic safety and enhances the driving experience.
Furthermore, V2V communication helps to increase the efficiency of transportation systems via platooning \cite{jnl:zeng18}.
\end{comment}
However, the performance of autonomous applications such as real-time navigation, collision avoidance, and platooning heavily rely on the ability to communicate with extremely low errors and delays. 
In this regard, achieving ultra-reliable low-latency communication (URLLC) is instrumental in enabling this vision~\cite{tech:5gcar}.
Since over-the-air latency and queuing latency are coupled, 
ensuring low queuing latency is required to achieve the target end-to-end latency of 1\,ms calling for efficient radio resource management (RRM) techniques \cite{tech:3gpp,jnl:mozaffari16,pap:ikram17}.
Towards this end, the majority of existing works including \cite{pap:ikram16,jnl:TLiu17} and \cite{jnl:mei18} focus on improving the average latency of vehicular networks while a handful such as \cite{pap:ikram17} and \cite{jnl:liu18} impose a probabilistic constraint to maintain small queue lengths.

Although a probabilistic constraint on the queue length improves network reliability, it fails to control rare events in which large queue lengths occur with low probability.
Therefore, some of the vehicular users (VUEs) may experience unacceptable latencies yielding degraded performance \cite{jnl:bennis18}.
\emph{Extreme value theory} (EVT), a powerful tool that characterizes the occurrences of extreme events with low probability, can be used to model the tail distribution in terms of three parameters known as location, scale, and shape~\cite{book:EVTfinkenstad03}. 
EVT asymptotically characterizes the statistics of extreme events by providing analytical models to analyze network traffic, worst-case delays, peak-to-average-power ratios, and ultra-reliable V2V communication in wireless systems \cite{jnl:liu18,pap:liu07,jnl:MOURADIAN16,pap:wei02}.
The challenge behind using EVT lies in gathering sufficient samples of extreme events that occur rarely to accurately model the tail distributions.
In V2V communication, VUEs may have access to limited number of queue length samples locally exceeding a high threshold, and hence they are unable to estimate the tail distribution of network-wide queue lengths.
Therefore, roadside units (RSUs) can assist in gathering samples over the network at the cost of additional data exchange overheads.
Furthermore, VUEs may be unwilling to share their individual queue state information (QSI) with an RSU and other VUEs as it may exhaust resources available for V2V communication.

This shortcoming warrants a collaborative learning model that does not rely on sharing individual QSI.
Recently, \emph{federated learning} (FL) was proposed as a decentralized learning technique where training datasets are unevenly distributed over learners, instead of centralizing all the data \cite{jnl:jakub16}.  
FL allows building learning models by sharing local models (set of learning parameters) with a central entity within few communication iterations.
Another notable advantage of FL is that it does not rely on synchronization among learners (e.g., VUEs).
Henceforth, even during a loss of connectivity between VUEs and RSUs, VUEs can still build their local models and navigate; this feature is crucial for highly dynamic and mission critical V2V communication.
To the best of our knowledge, no prior work has studied the use of federated learning in the context of ultra-reliable and low-latency communication.

The main contribution of this paper is to propose a distributed joint transmit power and resource allocation framework for enabling ultra-reliable and low-latency vehicular communication.
We formulate a network wide power minimization problem while ensuring low latency and high reliability in terms of probabilistic queue lengths. 
To model reliability, we first obtain the statistics of the queue lengths exceeding a high threshold by using the EVT concepts of a \emph{generalized Pareto distribution} (GPD)~\cite{book:EVTfinkenstad03}.
Using the statistics of the GPD, we impose a local constraint on the excess queues for each VUE.
Here, the parameters of the GPD known as scale and shape, are obtained by using maximum likelihood estimation (MLE).
In contrast to the classical design of MLE which requires a central controller (e.g., RSU) collecting samples of queues exceeding a threshold from all VUEs in the network,
leveraging principles of FL, we propose a distributed MLE technique that does not require sharing the queue length samples with an RSU or VUEs.
Specially, over time, every vehicle builds and sends its own local model (GPD parameters) to the RSU, which in turn aggregates, does model averaging across vehicles, and feeds back the global model to VUEs.
Leveraging different time scales, each VUE learns its GPD parameters locally in a short time scale while the model averaging (global learning) takes place in a longer time scale. 
Finally, Lyapunov optimization is used to decouple and solve the network-wide optimization problem per VUE.
Simulation results show that the proposed approach estimates GPD parameters as accurate as a centralized learning module and yields significant gains in terms of reducing VUEs with large queue lengths and improving power consumption.

The rest of the paper is organized as follows.
Section \ref{sec:system_model} describes the system model and the problem formulation.
The distributed solution based on EVT and Lyapunov optimization is presented in Section \ref{sec:distributed_solution}.
In Section \ref{sec:federated_learning}, estimation of the extreme value distribution using FL is discussed.
Section \ref{sec:results} evaluates the proposed solution by extensive set of simulations.
Finally, conclusions are drawn in Section \ref{sec:conclusion}.
\section{System Model and Problem Definition}\label{sec:system_model}

\begin{figure}[!t]
	\centering
	\includegraphics[width=\myfigfactor\linewidth]{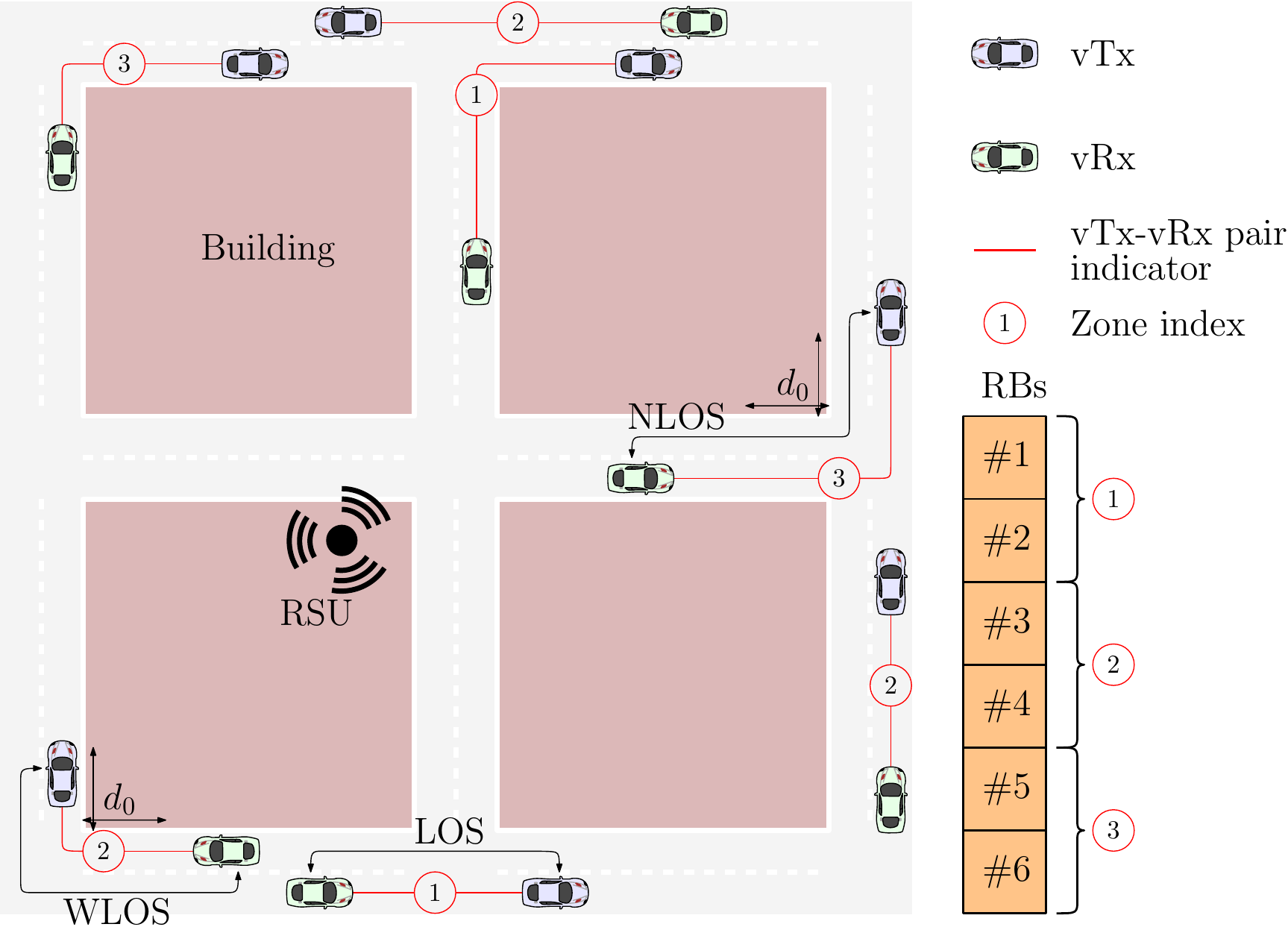}
	\caption{Simplified illustration of the system model containing \txr~pairs along their zone indexes and RB allocation over zones.}
	\label{fig:system_model}
\end{figure}

Consider a vehicular network consisting a set of $\vueSet$ of $\VUE$ VUE pairs exchanging information, with the aid of an RSU that allocates a set $\rbSet$ of resource blocks (RBs) over a partition of the network $\zoneSet$ defined as zones.
Here, a zone consists of VUE pairs that are located far from each other.
Moreover, RBs are orthogonally allocated over the zones to reduce the interference among nearby VUE pairs. 
Hence, a VUE transmitter-receiver (\txr) pair $\vue$ is only allowed to use the set $\rbSet_{\zone(t,\vue)}$ of RBs allocated for its corresponding zone $\zone(t,\vue)$ at time $t$.
The system layout is illustrated in Fig. \ref{fig:system_model}.

Let $\txpowerVec_{\vue}(t) = [\txpower_{\vue}^{\rb}(t)]_{\rb\in\rbSet_{\zone(t,\vue)}}$ and $\channelVec_{\vue\vue'}(t)=[\channel_{\vue\vue'}^{\rb}(t)]_{\rb\in\rbSet_{\zone(t,\vue)}}$ be, respectively, the transmit power of VUE $\vue$ and the channel gain between vTx $\vue$ and vRx $\vue'$ over the RBs at time $t$.
Depending on whether the vTx and vRx are located in the same lane or separately in perpendicular lanes, the channel model is categorized into three types:
\emph{i) Line-of-sight} (LOS): both vTx $\vue$ and vRx $\vue'$ are located in the same lane,
\emph{ii) Weak-line-of-sight}  (WLOS): vTx $\vue$ and vRx $\vue'$ are in perpendicular lanes and at least one of them is located at a distance of	 no more than $\distanceBound$ from the corresponding intersection,
and
\emph{iii) Non-line-of-sight}  (NLOS), otherwise.
Let $\coordinatesTx$ and $\coordinatesRx$ be the Cartesian coordinates of vTx $\vue$ and vRx $\vue'$, respectively.
As such, the path loss model of channel $\channel_{\vue'\vue}$ is based on the path loss model for urban areas using 5.9\,GHz carrier frequency
as follows~\cite{pap:mangel11}:
\begin{equation}\label{eqn:path_loss_model}
	\pathloss_{\vue\vue'} = 
	\begin{cases}
		\pathlossCoefficient \norm{\coordinatesTx-\coordinatesRx}{2}^{-\pathlossExponent} & \text{for LOS}, \\
		\pathlossCoefficient \norm{\coordinatesTx-\coordinatesRx}{1}^{-\pathlossExponent} & \text{for WLOS}, \\
		\pathlossCoefficient' ( |x_{\vue}-x_{\vue'}| \cdot |y_{\vue}-y_{\vue'}| )^{-\pathlossExponent} & \text{for NLOS},
	\end{cases}
\end{equation}
where $\norm{\vectx}{l}$ is the $l$-th norm of vector $\vectx$, $\pathlossExponent$ is the path loss exponent, and the path loss coefficients $\pathlossCoefficient$ and $\pathlossCoefficient'$ satisfy $\pathlossCoefficient' < \pathlossCoefficient (\frac{\distanceBound}{2})^{\pathlossExponent}$.
The transmission rate between the \txr~pair $\vue$ is given by,
\begin{equation}\label{eqn:rate_vue_pair}
	\rate_{\vue}(t) 
	= \textstyle \sum\limits_{\rb\in\rbSet_{\zone(t,\vue)}} \rate_{\vue}^{\rb}(t) 
	= \textstyle \sum_{\rb} \bandwidth 
	\log_2 \left( 1 + \frac{ \channel_{\vue\vue}^{\rb}(t) \txpower_{\vue}^{\rb}(t) }{ \interference_{\vue}^{\rb}(t) + \noise }\right),
\end{equation}
where $\interference_{\vue}^{\rb}(t) = \sum_{\vue'\in \vueSet\setminus\{\vue\} }  \channel_{\vue'\vue}^{\rb}(t) \txpower_{\vue'}^{\rb}(t)$ is the interference from other vTxs, $\bandwidth$ is the bandwidth of each RB, and $\noiseAlone$ is the noise power spectral density.
At each time $t$, $\arrival_{\vue}(t)$ data bits are randomly generated with a mean of $\arrivalAvg_{\vue}$ at vTx $\vue$ that must be delivered to its corresponding vRx.
Thus, at the vTx, a data queue is maintained which has the following dynamics:
\begin{equation}\label{eqn:queue_dynamics}
	\queue_{\vue}(t+1) = [\queue_{\vue}(t) + \arrival_{\vue}(t) - \rate_{\vue}(t)]^+,
\end{equation}
where $[x]^+ = \max(x,0)$.

Since most vehicular applications rely on exchanging information with guaranteed low end-to-end latency (mission critical), such networks demand higher network resources.
As such, 
our objective is to minimize the network-wide power consumption while ensuring URLLC.
Here, the reliability is achieved by ensuring queue stability for each vTx while maintaining outages in terms of exceeding a pre-defined queue threshold $\queueTH$ below a certain probability $\outage$. 
The reliability conditions are defined as follows:
\begin{gather}
	\label{eqn:queue_stability}
	\queueAvg{\vue} = \textstyle \lim_{\timeblock \to \infty} \frac{1}{\timeblock}\sum_{t=1}^{\timeblock} \queue_{\vue}(t) < \infty \qquad \forall \vue\in\vueSet, \\
	\label{eqn:queue_reliability}
	\probability( \queue_{\vue}(t) \geq \queueTH) \leq \outage \qquad \forall \vue\in\vueSet, \forall t.
\end{gather}

Note that the above reliability constraints have no control over the extreme cases with $\queue_{\vue}(t) > \queueTH$.
This impacts the worst case network queuing latency (as well as end-to-end latency \cite{tech:3gpp,jnl:mozaffari16,pap:ikram17}) and thus, needs to be properly addressed.
In this regard, let $\queuePoT(t)$ be a sample of excess value of the queue of any VUE over the threshold $\queueTH$ at time $t$ and $\queuePoT\in\{\queuePoT(t)\}_{\forall t}$.
By imposing a constraint,
\begin{equation}
\label{eqn:constraintNW_extremes_bounds}
	\textstyle \lim\limits_{\timeblock \to \infty} \sum_{t=1}^{\timeblock} \big( \queue_{\vue}(t) - \queueTH \big)\indictsimp{\queue_{\vue}(t)} / \sum_{t=1}^{\timeblock} \indictsimp{\queue_{\vue}(t)} \leq \expect[\queuePoT],
\end{equation}
each VUE $\vue$ can control its queue below the worst case queuing latency expected over the network.
Here, $\indictsimp{x}$ is an indicator function such that $\indictsimp{x}=1$ when $x>\queueTH$, and $\indictsimp{x}=0$ otherwise.
The network-wide power minimization problem will be:
\begin{subequations}\label{eqn:optimization_NW}
\begin{eqnarray}
	\label{eqn:objectiveNW_min_power}
	\underset{[\txpowerVec_{\vue}(t)]_{\forall\vue\in\vueSet}^{\forall t}}{\text{minimize}} && \textstyle \lim_{\timeblock \to \infty}  \frac{1}{\timeblock}\sum_{t=1}^{\timeblock}   \sum_{\vue\in\vueSet} \one\transpose \txpowerVec_{\vue}(t)  \\
	\label{eqn:constraintNW_reliability_modified}
	\text{subject to} && \eqref{eqn:queue_dynamics}\text{-}\eqref{eqn:constraintNW_extremes_bounds}, \\
	\label{eqn:constraintNW_control_domain}
	&& \txpowerVec_{\vue}(t) \succcurlyeq \zero, \quad \one\transpose\txpowerVec_{\vue}(t) \leq \txpowerMax \quad \forall \vue\in\vueSet.
\end{eqnarray}
\end{subequations}
Here, \eqref{eqn:constraintNW_reliability_modified} ensures queue dynamics and reliability while controlling the worst-case latency over all VUEs and $\txpowerMax$ is the transmit power budget of a VUE.
Solving \eqref{eqn:optimization_NW} which implies finding the optimal transmission control policy over time, is challenging due to two reasons:
\emph{i)} A decision at time $t$ relies on future network states,  
and
\emph{ii)} The characteristics of the distribution of $\queuePoT$ for constraint \eqref{eqn:constraintNW_extremes_bounds} are unavailable.
Furthermore, in the current context, solving  \eqref{eqn:optimization_NW} relies on a centralized controller.
This requires exchanging channel state information (CSI) and QSI over the whole network yielding unacceptable signaling overheads.
Therefore, new analytical tools are required to build a tractable model and to provide a distributed solution.

\section{Proposed Distributed Framework using EVT and Lyapunov Optimization}\label{sec:distributed_solution}

Proposing a distributed solution for network-wide power minimization problem relies on the ability to decouple \eqref{eqn:optimization_NW} over VUE pairs.
Therefore, the rest of the discussion investigates techniques to decouple the objective function \eqref{eqn:objectiveNW_min_power} and the constraint \eqref{eqn:constraintNW_extremes_bounds} based on the statistics of excess queues over the network.

\subsection{Modeling Excess Queues Using Extreme Value Theory}\label{sec:evt}

The excess queues $\{ \queuePoT(t) \}_{\forall t}$ are known as extreme statistics of the system,
and can be characterized using EVT.
Assume that the individual queues at a given time $[\queue_{\vue}(t)]_{\vue\in\vueSet}$ are samples of independent and identical distributions (i.i.d.) and the queue threshold $\queueTH$ is large.
Then, the distribution of $\queuePoT$ can be modeled as a generalized Pareto distribution (GPD) using \cite[Theorem 3.2.5]{book:EVTfinkenstad03}.
This theorem mainly shows that as $\queueTH \to \sup \{ q | \probability(\queuePoT(t) > q ) > 0 \}$, the conditional probability distribution of $\queuePoT(t) = \queue(t)- \queueTH$ is given by,
\begin{equation}\label{eqn:GPD}
\gpd^{\gpdCombined} (\queueMAXvar) = 
\begin{cases}
\frac{1}{\gpdScale}(1 + \gpdShape\queueMAXvar/\gpdScale)^{-1-1/\gpdShape} & \text{for~} \gpdShape \neq 0, \\
\frac{1}{\gpdScale} e^{-\queueMAXvar/\gpdScale}  & \text{for~} \gpdShape = 0,
\end{cases}
\end{equation}
with $\gpdCombined = [\gpdScale,\gpdShape]$, where $\gpdShape$ and $\gpdScale(>0)$ are the shape and scale parameters, respectively.
Here, $\queueMAXvar \geq 0$ if $\gpdShape \geq 0$ while $0 \leq \queueMAXvar \leq -\gpdScale/\gpdShape$ when $\gpdShape > 0$.
Furthermore, $\expect[\queuePoT]$ is bounded and equivalent to $\gpdScale / (1-\gpdShape)$, only if $\gpdShape < 1$.
In this regard, 
constraint \eqref{eqn:constraintNW_extremes_bounds} for all $\vue\in\vueSet$ can be rewritten as follows:
\begin{equation}\label{eqn:constraintNW_extremes_bounds_modified}
	\textstyle \lim\limits_{\timeblock \to \infty} \sum_{t=1}^{\timeblock} \big( \queue_{\vue}(t) - \queueTH \big)\indictsimp{\queue_{\vue}(t)} / \sum_{t=1}^{\timeblock} \indictsimp{\queue_{\vue}(t)} \leq \gpdScale / (1-\gpdShape).
\end{equation}
Assisted by the RSU, each VUE pair estimates $\gpdShape$ and $\gpdScale$ locally without sharing the QSI allowing to decouple the constraint \eqref{eqn:constraintNW_extremes_bounds} and impose it locally as in \eqref{eqn:constraintNW_extremes_bounds_modified}.

\subsection{Lyapunov Optimization}\label{sec:lyapunov}

By using EVT to model $\queuePoT$ and its expected value, we recast the original problem into an equivalent form:
\begin{subequations}\label{eqn:optimization_NW_modified}
	\begin{eqnarray}
	\underset{[\txpowerVec_{\vue}(t)]_{\forall\vue\in\vueSet}^{\forall t}}{\text{minimize}} && \textstyle \lim_{\timeblock \to \infty}  \frac{1}{\timeblock}\sum_{t=1}^{\timeblock} \left(  \sum_{\vue\in\vueSet} \one\transpose \txpowerVec_{\vue}(t) \right) \\
	\label{eqn:constraint_all}
	\text{subject to} && \eqref{eqn:queue_stability}, \eqref{eqn:queue_reliability}, \eqref{eqn:constraintNW_control_domain}, \eqref{eqn:constraintNW_extremes_bounds_modified}.
	\end{eqnarray}
\end{subequations} 

To present a tractable solution for the above modified stochastic optimization problem, we resort to Lyapunov optimization.
First, the time average constraints need to be modeled as virtual queues.
As such, the reliability constraint \eqref{eqn:queue_reliability} can be recast as $\expect[\indictsimp{\queue_{\vue}}] \leq \outage$ for each VUE $\vue\in\vueSet$.
The goal is to introduce a virtual queue for the aforementioned constraint instead of \eqref{eqn:queue_stability} and \eqref{eqn:queue_reliability}.
In order to scale this virtual queue with the actual queue size, both sides of the aforementioned constraint are scaled by the queues, and thus, it becomes $\expect[\indictsimp{\queue_{\vue}} \queue_{\vue}] \leq \outage \expect[\queue_{\vue}]$.
Now the time average constraints in \eqref{eqn:constraint_all} for all $\vue\in\vueSet$ are modeled by virtual queues as follows:
\begin{subequations}\label{eqn:virtual_queues_for_constraints}
	\begin{align}
	&\vqQ_{\vue}(t+1) = [\vqQ_{\vue}(t) + ( \indictsimp{\queue_{\vue}(t)} - \outage ) \queue_{\vue}(t+1) ]^+, \\
	&\vqEXq_{\vue}(t+1) = [\vqEXq_{\vue}(t) + \big( \queue_{\vue}(t+1) - \queueTH - \textstyle \frac{\gpdScale}{1-\gpdShape} \big) \indictsimp{\queue_{\vue}(t)} ]^+. 
	\end{align}
\end{subequations}

Let $\queueCombined_{\vue}(t)=[\queue_{\vue}(t),\vqQ_{\vue}(t),\vqEXq_{\vue}(t))]$ be the combined queue with $\vect{\queueCombined}(t) = [\queueCombined_{\vue}(t)]_{\vue\in\vueSet}$ and its quadratic Lyapunov function $\lyapunov{\vect{\queueCombined}(t)}=\frac{1}{2}\vect{\queueCombined}\transpose(t)\vect{\queueCombined}(t)$.
The one-slot drift of the Lyapunov function is defined as $\lyapunovDrift = \lyapunov{\vect{\queueCombined}(t+1)} - \lyapunov{\vect{\queueCombined}(t)}$.
Given that,
\begin{equation*}
([\queue + \arrival - \rate]^+)^2 \leq \queue^2 + (\arrival - \rate)^2 + 2\queue(\arrival - \rate), 
\end{equation*}
the Lyapunov drift can be simplified and bounded as follows:
\begin{multline}\label{eqn:lyapunov_drift}
\lyapunovDrift \leq \lyapunovBound + \textstyle \sum\limits_{\vue\in\vueSet} \Big[
\lyapunovConst 
 + \big( \arrival_{\vue}(t) - \rate_{\vue}(t) \big) \Big\{ 
(1 + \outage^2)\queue_{\vue}(t) 
- \outage \vqQ_{\vue}(t) \\
+ [  
2(1-\outage) \queue_{\vue}(t) + \vqQ_{\vue}(t) +
\vqEXq_{\vue}(t) - \queueTH - \textstyle \frac{\gpdScale}{1-\gpdShape} 
] \indictsimp{\queue_{\vue}(t)} 
\Big\}
\Big],
\end{multline}
where $\lyapunovBound = \sum_{\vue} \frac{1}{2} \big\{
(1 + \indictsimp{\queue_{\vue}(t)} \big( \queue_{\vue}(t)^2 + \big( \arrival_{\vue}(t) - \rate_{\vue}(t) \big)^2 \big) )
+ \indictsimp{\queue_{\vue}(t)} (\queueTH + \frac{\gpdScale}{1-\gpdShape} )^2
\big\}$ is a uniform bound of the Lyapunov drift, and
$\lyapunovConst = \queue_{\vue}(t) \big\{ [ \vqQ_{\vue}(t) + \vqEXq_{\vue}(t) - \queueTH - \textstyle \frac{\gpdScale}{1-\gpdShape} ] \indictsimp{\queue_{\vue}(t)} - \outage \vqQ_{\vue}(t)
\big\}
- \indictsimp{\queue_{\vue}(t)} \vqEXq_{\vue}(t) (\queueTH + \textstyle \frac{\gpdScale}{1-\gpdShape})$ is a fixed term for VUE $\vue$ at time $t$ which is independent of the control parameters.

The conditional expected Lyapunov drift at time $t$ is defined as $\expect[\lyapunov{\vect{\queueCombined}(t+1)} - \lyapunov{\vect{\queueCombined}(t)} |\vect{\queueCombined}(t) ]$.
Let $\lyapunovTradeoff \geq 0$ be a parameter which controls the tradeoff between queue length and the accuracy of the optimal solution of \eqref{eqn:optimization_NW_modified}.
Introducing a penalty term $\lyapunovTradeoff \expect[\sum_{\vue} \one\transpose \txpowerVec_{\vue}|\vect{\queueCombined}(t)]$ to the expected drift and minimizing the upper bound of the drift plus penalty (DPP),
$ \lyapunovTradeoff \expect[\sum_{\vue} \one\transpose \txpowerVec_{\vue}|\vect{\queueCombined}(t)] + \expect[\lyapunovDrift|\vect{\queueCombined}(t)] $,
yields the network control policies.
Thus, the objective is to minimize the following upper bound:
\begin{multline}
	\sum_{\vue\in\vueSet} 
	\lyapunovTradeoff \one\transpose \txpowerVec_{\vue}
	+ \lyapunovConst
	+ \big( \arrival_{\vue}(t) - \rate_{\vue}(t) \big) \Big\{ 
	(1 + \outage^2)\queue_{\vue}(t) 
	- \outage \vqQ_{\vue}(t) \\
	+ [  
	2(1-\outage) \queue_{\vue}(t) + \vqQ_{\vue}(t) +
	\vqEXq_{\vue}(t) - \queueTH - \textstyle \frac{\gpdScale}{1-\gpdShape} 
	] \indictsimp{\queue_{\vue}(t)} 
	\Big\},
\end{multline}
at each time $t$.
Assuming that VUEs maintain channel-quality indicators (CQIs), each VUE can estimate the interference $\interference_{\vue}^{\rb}(t) \simeq \interferenceEST_{\vue}^{\rb}(t)$ based on its past experiences (time average)~\cite{pap:luoto17}.
Hence, the minimization of the above upper bound can be decoupled among VUEs as follows:
\begin{subequations}\label{eqn:optimization_power_vue}
	\begin{alignat}{2}
	\underset{\txpowerVec_{\vue}(t)}{\text{minimize}} &  \textstyle \sum\limits_{\rb\in\rbSet_{\zone(t,\vue)}} \!\!\!\! \big[ \lyapunovTradeoff\txpower_{\vue}^{\rb}(t)
	- \si_{\vue}(t) \ln \big( 1 + \cinr_{\vue}^{\rb}(t) \txpower_{\vue}^{\rb}(t) \big) \big] \\
	\label{eqn:constraint_sum_power} \text{subject to} & \quad \textstyle  \sum_{\rb\in\rbSet_{\zone(t,\vue)}} \txpower_{\vue}^{\rb}(t) \leq \txpowerMax, \\
	& \quad \txpower_{\vue}^{\rb}(t) \geq 0 \qquad \forall \rb\in\rbSet_{\zone(t,\vue)},
	\end{alignat}
\end{subequations} 
where $\si_{\vue}(t) = \frac{\bandwidth}{\ln 2} \big\{ (1 + \outage^2)\queue_{\vue}(t) 
- \outage \vqQ_{\vue}(t) 
+ [  
2(1-\outage) \queue_{\vue}(t) + \vqQ_{\vue}(t) +
\vqEXq_{\vue}(t) - \queueTH - \frac{\gpdScale}{1-\gpdShape} 
] \indictsimp{\queue_{\vue}(t)} 
\big\}$ and $\cinr_{\vue}^{\rb}(t) = \frac{ \channel_{\vue\vue}^{\rb}(t)  }{ \interferenceEST_{\vue}^{\rb}(t) + \noise }$.
The optimal solution of the convex optimization problem of \eqref{eqn:optimization_power_vue} is obtained by the \emph{water-filling algorithm} such that  $[\txpower_{\vue}^{\rb}(t)]\optimal = [\frac{\si_{\vue}(t)}{\lyapunovTradeoff + \dualPower\optimal_{\vue}(t)} - \frac{1}{\cinr_{\vue}^{\rb}(t)}]^+$, where $\dualPower_{\vue}(t) \geq 0$ is the Lagrangian dual coefficient corresponding to the constraint \eqref{eqn:constraint_sum_power}.

\section{Learning the Parameters of the Maximum Queue Distribution}\label{sec:federated_learning}

\begin{figure}[!t]
	\centering
	\includegraphics[width=\myfigfactor\linewidth]{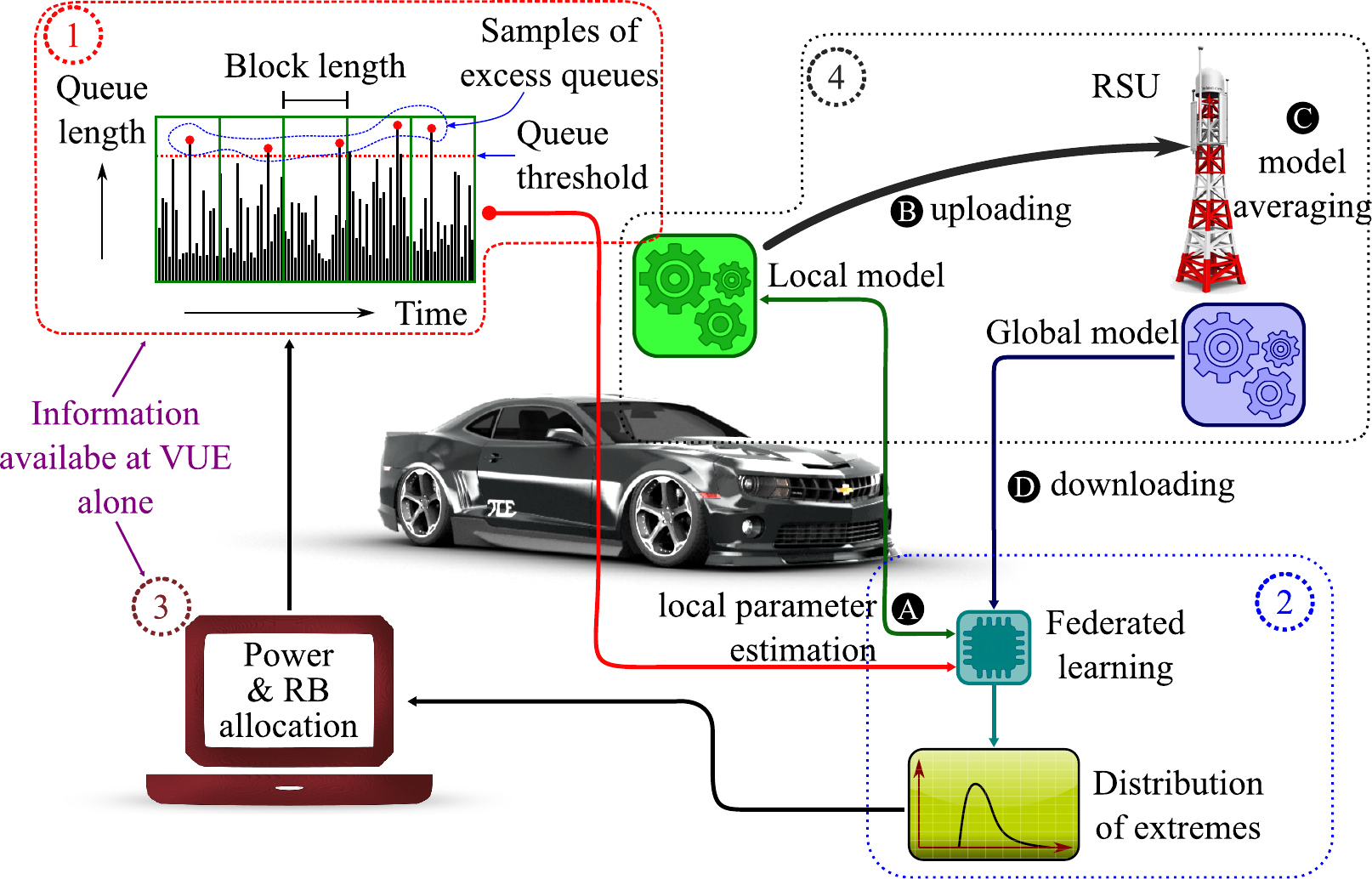}
	\caption{Interrelationships of the processes between VUEs and RSU: 1) excess queue sampling, 2) GPD parameter estimation, 3) transmit power and RB allocation, and 4) local and global models exchange with the RSU.}
	\label{fig:processes_of_vue}
\end{figure}

The proposed distributed control mechanism relies on the characteristics of the excess queue distribution $\gpd^{\gpdCombined} (\queueMAXvar)$.
Therefore, the estimation of the parameters $\gpdScale$ and $\gpdShape$  with high accuracy using QSI samples gathered at each VUE is crucial. 
In this regard, modeling the excess queue distribution requires a central controller (e.g., the RSU) to compute and communicate with all VUEs at each time $t$.
However, this RSU-centric approach is impractical due to the fact that:
\emph{i)} The overhead needed for frequent communications with all the VUEs in a highly dynamic network will degrade the network-wide performance, and
\emph{ii)} VUEs may prefer not to share their QSI with other vehicles, in which warrants collaborative learning techniques \cite{jnl:jakub16,pap:hamidouche18}.
Therefore, next, we propose a distributed solution based on FL that allows each VUE to learn the GPD parameters (\emph{local model}) individually using local QSI observations and minimal communication with the RSU.
In turn, the RSU averages the received parameters (\emph{global model}) and sends them back to the VUEs, as summarized in Fig. \ref{fig:processes_of_vue}.

\subsection{Queue sampling via block maxima (BM) }

Let $\blocklength$ be the block length (or time window) during which each VUE draws a maximum queue length sample.
The size of $\blocklength$ should be sufficiently large to minimize correlation between QSI samples while being sufficiently small to reduce the sampling overhead process.
Since each VUE can independently perform QSI sampling process, the total number of samples may vary across VUEs. 
In this regard, the $\block$-th sample of VUE $\vue$ is $\sample_{\vue}^{\block} = \max_{t\in\blockTimeSet_{\block}} \big( \queue_{\vue}(t) - \queueTH  \big) \indictsimp{\queue_{\vue}(t)}$  where $\blockTimeSet_{\block} = \{ (\block-1)\blocklength, (\block-1)\blocklength+1, \dots, \block\blocklength-1 \}$ and the set of samples with $\BLOCK_{\vue}$ is $\sampleSet_{\vue}= \{\sample_{\vue}^{\block}\}_{\block\in\{\seta{\BLOCK_{\vue}}\}}$.
The QSI sampling procedure is illustrated in Fig. \ref{fig:processes_of_vue}.

\subsection{Estimating the GPD parameters}

As shown in Section \ref{sec:evt}, the distribution of the excess queue length samples is characterized by two parameters $\gpdCombined = [\gpdScale,\gpdShape]$ which need to be estimated.
For this purpose, we use MLE~\cite{book:paul84}.
The goal of MLE is to find the best set of parameters $\gpdCombined$ that fits the GPD $\gpdML(\cdot)$ to the samples via maximizing the log likelihood function, or minimizing the negative of it, as follows:
\begin{equation}\label{eqn:MLE_global}
	\underset{\gpdCombined\in\gpdCombinedFeasible(\sampleSet)}{\text{minimize}} \quad  \loglikelihood(\sampleSet)
	= -\frac{1}{\setSize{\sampleSet}}\sum_{\sample\in\sampleSet} \log \gpdML(\sample),
\end{equation} 
where $\gpdCombinedFeasible(\sampleSet) = \{ [\gpdScale,\gpdShape]\in\Re^{2} | \gpdScale > 0, \gpdShape < 1, 1 + \gpdShape \sample /\gpdScale) \geq 0 \text{~for all~} \sample\in\sampleSet \}$ is the feasible set  and  $\sampleSet = \{\sampleSet_{\vue}\}_{\vue\in\vueSet}$ is the set of network queue length samples.
Ideally, solving \eqref{eqn:MLE_global} requires a central processor (RSU in our scenario) with queue length samples of all the VUEs.
Note that the likelihood function is a smooth function of $\gpdCombined$ and a summation over all the samples in $\sampleSet$, and thus, its gradient over a sample $\sample$ is given as follows:

\begin{proposition}\label{prop:derivative_coefficient}
	The derivative coefficient of the negative log-likelihood function of GPD at queue length sample $\sample$ w.r.t. $\gpdCombined$ is,
	\begin{equation}\label{eqn:gradients}
	\grad{\gpdCombined}{\loglikelihood(\sample)}
	= \begin{bmatrix}
	\daba{\loglikelihood(\sample)}{\gpdScale} \\
	\daba{\loglikelihood(\sample)}{\gpdShape}
	\end{bmatrix}
	= \begin{bmatrix}
	\frac{1}{\gpdScale} \big( \frac{1 + 1/\gpdShape}{ 1 + \gpdShape \sample / \gpdScale } - \frac{1}{\gpdShape} \big)\\
	\frac{(1 + 1/\gpdShape) ( 2 + \gpdShape \sample / \gpdScale )}{ 1 + \gpdShape \sample / \gpdScale } - \frac{ \ln (1 + \gpdShape \sample / \gpdScale) }{\gpdShape^2} 
	\end{bmatrix}.
	\end{equation}
\end{proposition}
\begin{IEEEproof}
	see Appendix \ref{appndx:gevd_gradient}.
\end{IEEEproof}

Therefore, any gradient descent technique-based algorithm can be used to determine the optimal $\gpdCombined\optimal$ in an iterative manner.
However, in our model, this is not applicable due to; \emph{i)} Presence of large number of VUEs and thus, increased overhead of sharing large amount of samples, and \emph{ii)} VUE unwillingness for sharing their QSI by exhausting network resources that may incur additional latencies.
Therefore, a mechanism to estimate the parameters locally while leveraging the RSU for model aggregation is needed.
For this purpose, we adopt FL as discussed below.

To this end, we rewrite the likelihood function as follows:
\begin{align}\label{eqn:likelihood_local}
	\loglikelihood(\sampleSet)
	= \frac{1}{\setSize{\sampleSet}}\sum_{\sample\in\sampleSet} \log \gpdML(\sample)
	= \sum_{\vue\in\vueSet}\sampleSizeRatio \loglikelihood(\sampleSet_{\vue}),
\end{align}
where $\sampleSizeRatio = \frac{\setSize{\sampleSet_{\vue}}}{\setSize{\sampleSet}} = \frac{\BLOCK_{\vue}}{\sum_{\vue'} \BLOCK_{\vue'}}$.
Here, the likelihood function of the network is presented as a weighted sum of likelihood functions per VUE.
Hereinafter, for simplicity, we use $\loglikelihood$ and $\loglikelihood_{\vue}$ instead of $\loglikelihood(\sampleSet)$ and $\loglikelihood(\sampleSet_{\vue})$, respectively.
The idea behind FL is to use $\loglikelihood_{\vue}$ to evaluate $\grad{\gpdCombined}{\loglikelihood_{\vue}}$ and $\gpdCombined$ locally, say $\gpdCombined_{\vue}$, and update the local estimations via sharing the individual learning \emph{models} $( \grad{\gpdCombined}{\loglikelihood_{\vue}}, \gpdCombined_{\vue}, \BLOCK_{\vue} )$.
As the local sample size can be large, the complexity of computing the gradient can be high. 
Therefore, it is assumed that each VUE uses the \emph{stochastic variance reduced gradient} (SVRG) to evaluate the gradients with a predefined step size $\stepsize(>0)$ \cite{jnl:jakub16}.
The GPD parameter estimation procedure using FL-based MLE is presented in Algorithm \ref{alg:federated_mle}.

\begin{algorithm}[!t]
	\caption{MLE for GPD using FL}
	\label{alg:federated_mle}
	\begin{algorithmic}[1]                    
		\STATE \textbf{input:} Gradients $\{\grad{\gpdCombined}{\loglikelihood_{\vue}}(0)\}_{\vue\in\vueSet}$, local estimations $\{\gpdCombined_{\vue}(0)\}_{\vue\in\vueSet}$, and step size $\stepsize$.
		\FOR{$\federatedTime = 1,2,\ldots$}
			\STATE Update $\gpdCombined(\federatedTime) = \gpdCombined(\federatedTime-1) + \sum_{\vue} \sampleSizeRatio \big( \gpdCombined_{\vue}(\federatedTime-1) - \gpdCombined(\federatedTime-1)\big)$.
			\STATE Compute $\grad{\gpdCombined}{\loglikelihood}(\federatedTime) = \frac{1}{\sum_{\vue'} \BLOCK_{\vue'}} \sum_{\vue} \grad{\gpdCombined}{\loglikelihood_{\vue}}(\federatedTime)$.
			\STATE Download the model $\big( \grad{\gpdCombined}{\loglikelihood}(\federatedTime), \gpdCombined(\federatedTime), \sum_{\vue} \BLOCK_{\vue}  \big)$ to all VUEs $\vueSet$.
			\FOR [in parallel] {each VUE $\vue\in\vueSet$} 
				\STATE \textbf{set:} $\gpdCombined_{\vue}(\federatedTime) = \gpdCombined(\federatedTime)$ and $\stepsize_{\vue} = \stepsize / \BLOCK_{\vue}$.
				\STATE Let $\{i^{\block}_{\vue}\}_{\block=1}^{\BLOCK_{\vue}}$ be a random permutation of $\sampleSet_{\vue}$.
				\FOR {$\block = \seta{\BLOCK_{\vue}}$}
					\STATE Compute $\vect{y}_{\vue} = \gpdCombined_{\vue}(\federatedTime) - \stepsize_{\vue} \big[ \grad{\gpdCombined}{\loglikelihoodSP{\gpdCombined_{\vue}(\federatedTime)}_{\vue}}(i_{\vue}^{\block}) - \grad{\gpdCombined}{\loglikelihoodSP{\gpdCombined(\federatedTime)}_{\vue}}(i_{\vue}^{\block}) + \grad{\gpdCombined}{\loglikelihood}(\federatedTime) \big]  $.
					\STATE Update $\gpdCombined_{\vue}(\federatedTime) = \argmin_{\gpdCombined_{\vue}\in\gpdCombinedFeasible(\sampleSet_{\vue})} \| \vect{y}_{\vue} - \gpdCombined_{\vue}  \|$.
				\ENDFOR
				\STATE Upload the model $\big( \grad{\gpdCombined}{\loglikelihood_{\vue}}(\federatedTime), \gpdCombined_{\vue}(\federatedTime), \BLOCK_{\vue}  \big)$ to RSU.
			\ENDFOR
		\ENDFOR
	\end{algorithmic}
\end{algorithm}

\section{Numerical Results}\label{sec:results}

In order to evaluate the performance of the proposed solution, a network based on a 250\,m$\times$250\,m Manhattan mobility model with nine intersections is considered.
Therein, a road consists of two lanes with 4\,m width for each direction.
VUE pairs are uniformly located within each lane with a fixed gap of 50\,m and vRx always follows vTx with 60\,kmph speed.
VUEs share 60 RBs and have a maximum transmit power of $\txpowerMax = 10$\,W.
The RB allocation per zone is adopted from \cite{pap:ikram17} and \cite{jnl:liu18}.
The rest of the parameter values are presented in Table~\ref{tab:simulation_parameters}.

\begin{table}
	\centering
	\caption{Simulation parameters \cite{pap:ikram16,pap:ikram17,jnl:liu18,pap:mangel11}.}
	\label{tab:simulation_parameters}
	\begin{tabular}{|c c|| c c|| c c|}
		\hline 
		Para. &  Value & Para. &  Value & Para. &  Value\\ 
		\hline \hline
		$\pathlossCoefficient$ & -68.5\,dBm & $\bandwidth$ & 180\,kHz& $\blocklength$ & 10 \\
		$\pathlossCoefficient'$ & -54.5\,dBm & $\noiseAlone$ & -174\,dBm/Hz & $\stepsize$ & (50,10)\\
		$\pathlossExponent$ & 1.61 & $\queueTH$  & 34.72 kb & $\grad{\gpdCombined}{\loglikelihood_{\vue}}(0)$ & (1,1000)\\
		$\distanceBound $ & 15\,m & $\outage$ & 0.001 & $\gpdCombined_{\vue}(0)$ & (50,0)\\
		\hline 
	\end{tabular} 
\end{table}

\subsection{Centralized vs distributed GPD parameter estimation}

\begin{figure}[!t]
	\centering
	\includegraphics[width=\myfigfactor\linewidth]{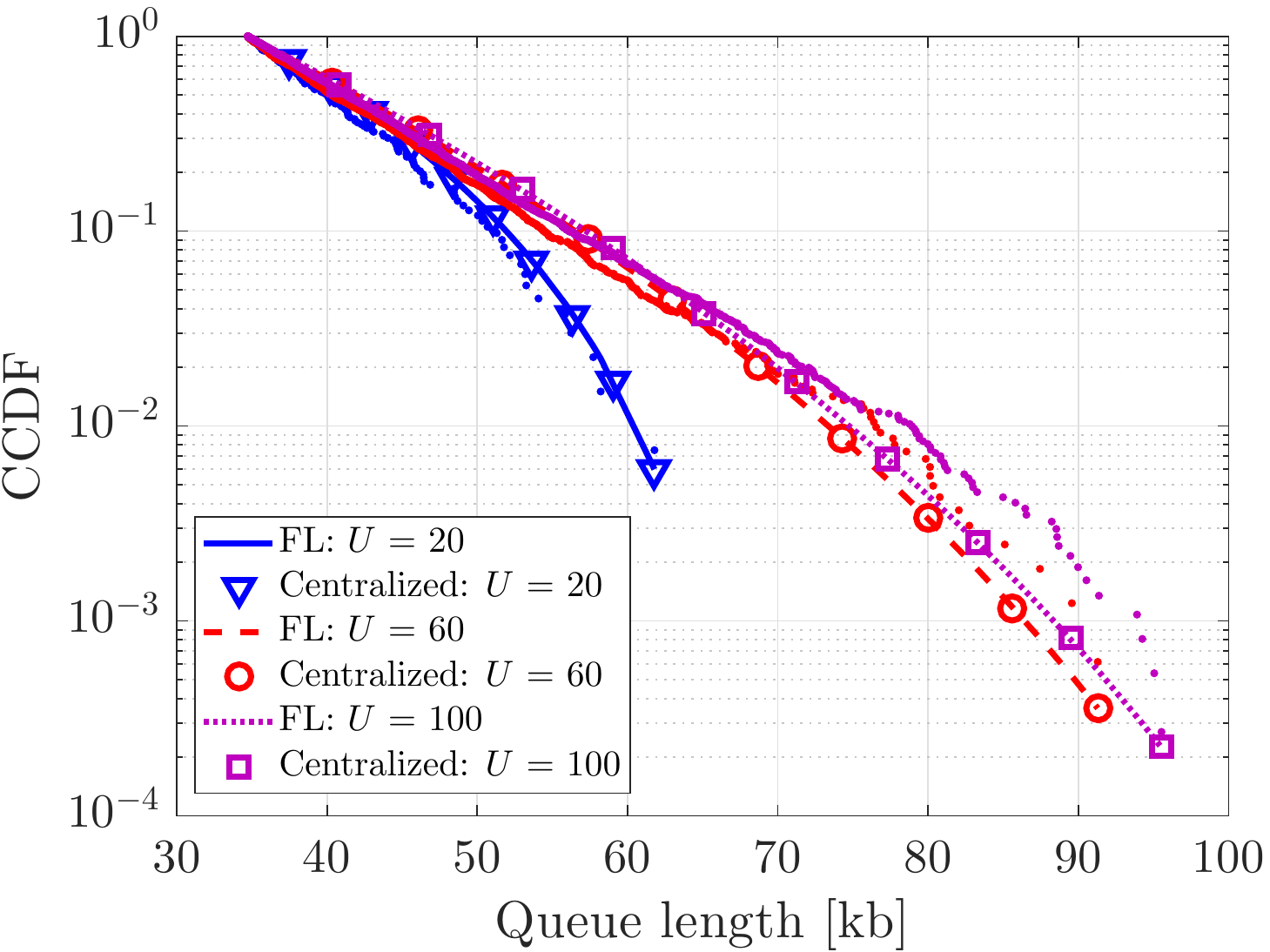}
	\caption{Complementary cumulative distribution functions (CCDF) of queue length using the proposed FL approach and the centralized SVRG method for different VUE densities.}
	\label{fig:test_evt}
\end{figure}

Here, the proposed FL approach is compared with a \emph{centralized} solution based on the SVRG method at the RSU to estimate the GPD parameters. 
Therein, all VUEs upload their local excess queue length samples to the RSU which estimates and shares the GPD parameters with all VUEs.
Fig. \ref{fig:test_evt} illustrates the estimated GPDs for both FL and centralized approaches with the corresponding excess queue length samples for different VUEs, $\VUE=$ 20, 60, and 100.
It can be noted that the FL estimations are almost equivalent to the centralized estimations.

\begin{figure}[!t]
	\centering
	\includegraphics[width=\myfigfactor\linewidth]{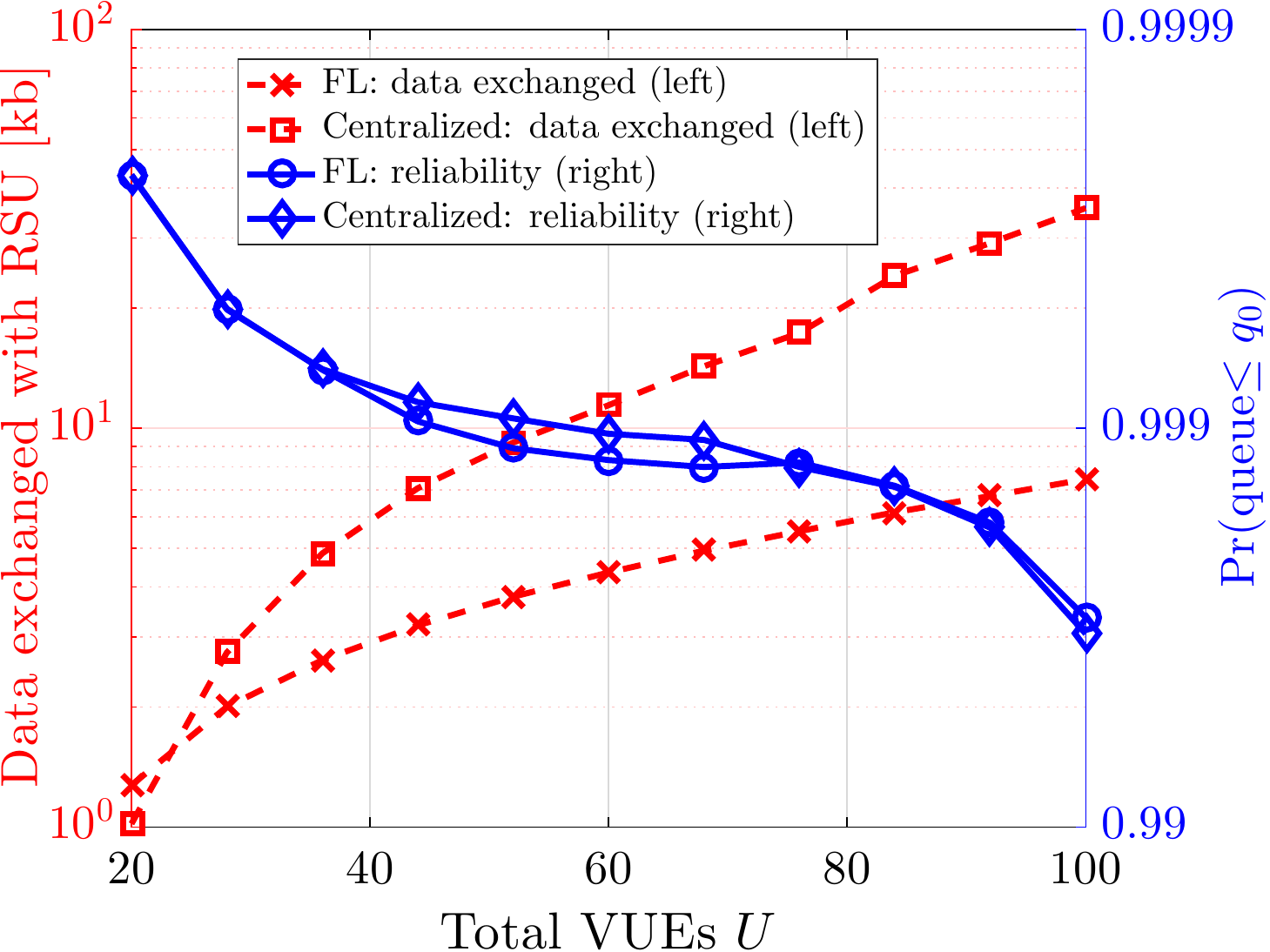}
	\caption{Comparison of the amount of data exchanged between RSU and VUEs (left) and the achieved reliability (right) for different VUE settings with two approaches used to estimate GPD parameters: the proposed FL and the centralized SVRG method.}
	\label{fig:FLvsCEN}
\end{figure}

In Fig. \ref{fig:FLvsCEN}, we compare the amount of data exchange and the achieved reliability in terms of maintaining the queue length below the threshold $\queueTH$ for different VUE densities.
As the reliability decreases with increasing the number of VUEs, our FL-based approach achieves a reliability that is nearly equal to the one resulting from the centralized approach.
Note that the centralized method requires all VUEs to upload all their queue length samples to the RSU and to receive the estimated GPD parameters.
In contrast, in the proposed method, VUEs upload their locally estimated model (GPD parameters, gradient, and sample size) and receive the global estimation of the model.
For fewer number of VUEs, $\VUE=20$, the sample size of the network is small, and thus the centralized method can operate efficiently using very few data samples. 
In contrast, in FL, VUEs must upload and download both parameters and gradients yielding higher data exchange compared to the centralized method.
However, as the number of VUEs increases (beyond 28), the sample size grows, and thus, the centralized method incurs higher amount of data exchanged between the RSU and VUEs compared to the proposed method.
The reductions of the exchanged data in the proposed method compared to the centralized model is about 27\% for $\VUE=28$ and improves up to 79\% when $\VUE=100$.
Since the overall performance in terms of reliability of both methods is similar, the proposed method becomes efficient due to the reductions in data exchange overhead.

\subsection{Performance evaluation}

\begin{figure}[!t]
	\centering
	\includegraphics[width=\myfigfactor\linewidth]{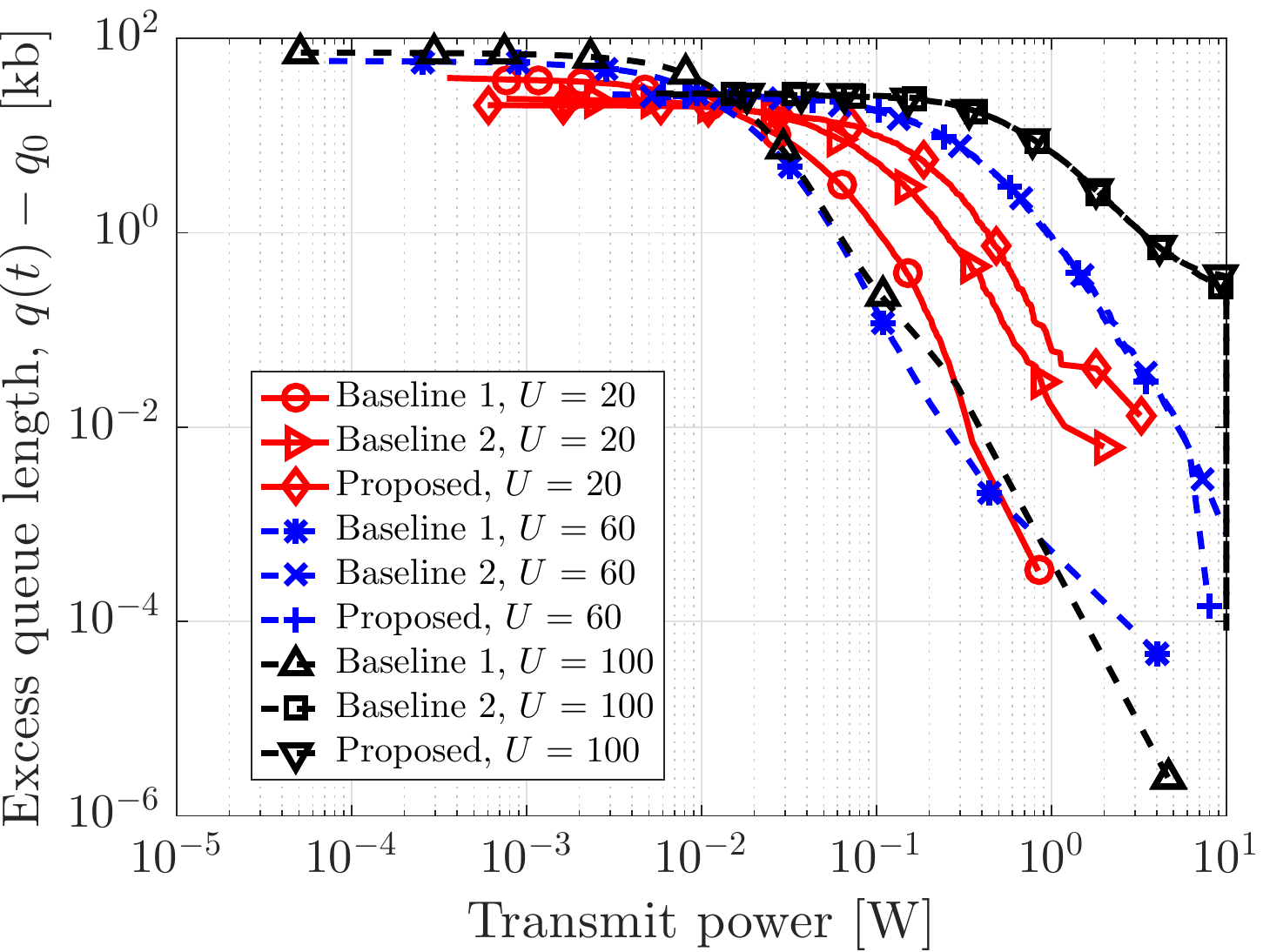}
	\caption{Transmit power versus excess queue length tradeoff for worst-case VUEs, i.e., when VUEs queue lengths become larger than $\queueTH$.}
	\label{fig:tradeoff}
\end{figure}

\begin{table}
	\centering
	\caption{Average power, queuing latency, excess queue lengths, and unreliability for Fig. \ref{fig:tradeoff} scenarios.}
	\label{tab:tradeoff_table}
	\begin{tabular}{| l | c | c | c | c | c | c |}
		\hline
		\multicolumn{1}{|c|}{Model} & \multicolumn{3}{|c|}{Avg. power [W] } & \multicolumn{3}{|c|}{Avg. latency [ms]} \\ \cline{2-7}
		\multicolumn{1}{|c|}{with $U$ =} & 20    & 60    & 100   & 20   & 60   & 100 \\ \hline
		Fixed power*                     & 0.1   & 0.1   & 0.1   & 2.58 & 2.36 & 3.37 \\
		Baseline 1                       & 0.007 & 0.008 & 0.011 & 1.54 & 5.24 & 11.02 \\
		Baseline 2                       & 0.013 & 0.065 & 0.242 & 0.92 & 0.39 & 0.41    \\
		Proposed                         & 0.013 & 0.058 & 0.242 & 1.15 & 0.48 & 0.4 \\ \hline \hline
		\multicolumn{1}{|c|}{Model} & \multicolumn{3}{|c|}{ Pr($\queue > \queueTH$) [$\times 10^{-3}$] } & \multicolumn{3}{|c|}{Avg. excess queue [kb]} \\ \cline{2-7}
		\multicolumn{1}{|c|}{with $U$ =} & 20    & 60    & 100  & 20   & 60   & 100 \\ \hline
		Fixed power*                     & 24.6  & 54.9  & 92.7 & 72.8 & 75.1 & 76.4 \\
		Baseline 1                       & 3.1   & 34.1  & 88.9 & 42.9 & 51.9 & 63.8 \\
		Baseline 2                       & 0.5   & 1.2  & 2.3   & 41.8 & 43.8 & 45.9    \\
		Proposed                         & 0.2   & 1.1  & 2.1   & 41.9 & 43.4 & 44.7 \\ \hline
		\multicolumn{7}{l}{* Not shown in Fig. \ref{fig:tradeoff} due to large excess queue lengths.}
	\end{tabular} 
\end{table}

Next, the proposed approach is compared with a \emph{``fixed power''} model using fixed transmit power and two other baseline models focusing on minimizing the total power consumption namely: 
\emph{i) \textbf{Baseline 1}}:
a V2V network with the only interest of queue stability \eqref{eqn:queue_dynamics}-\eqref{eqn:queue_stability},
and
\emph{ii) \textbf{Baseline 2}}:
a V2V network with the probabilistic constraint on average queue length \eqref{eqn:queue_dynamics}-\eqref{eqn:queue_reliability}.

Fig. \ref{fig:tradeoff} compares the transmit power and excess queue length tradeoff for VUEs exceeding the queue threshold $\queueTH$ for the aforementioned methods.
In Table \ref{tab:tradeoff_table}, the statistics of average transmit power, queuing latency, excess queue length, and unreliability defined as the probability that the queue length grows larger than $\queueTH$ corresponding to the scenarios used in Fig. \ref{fig:tradeoff} are presented.
Note that the fixed power method has no control on queue stability, and hence, it yields VUEs with large queues.
Therefore, the excess queues for the fixed power model is intentionally neglected from Fig. \ref{fig:tradeoff}.
From Table \ref{tab:tradeoff_table}, we can see that the fixed power model yields the highest excess queue lengths on average, as well as the highest number of VUEs exceeding $\queueTH$ for all  three cases with different VUE densities.

From Fig. \ref{fig:tradeoff} and Table \ref{tab:tradeoff_table}, we can see that Baseline 1 uses the lowest transmit powers in all three cases.
The reason is that Baseline 1's only concern is to minimize the transmit power while ensuring queue stability without a constraint on queue lengths.
However, Baseline 1 results in larger excess queue lengths as shown in Fig. \ref{fig:tradeoff}.
Moreover, Table \ref{tab:tradeoff_table} shows that Baseline 1 exhibits higher excess queue lengths on average, increased average latencies, and more VUEs exceeding $\queueTH$ compared to Baseline 2 and the proposed method.
It can be noted that this performance gap increases with the increasing VUE densities.

The advantage of Baseline 2 over Baseline 1 is in terms of the reduced excess queue lengths as shown in Fig. \ref{fig:tradeoff}.
Its improvements are close to the proposed method due to the additional probabilistic constraint on the queue length compared to Baseline 1.
Although the difference between the proposed approach and Baseline 2 method in Fig. \ref{fig:tradeoff} is hard to distinguish as $\VUE$ increases, from Table \ref{tab:tradeoff_table}, we can observe that the reductions of excess queue lengths on average increase up to 3\% in the proposed method over Baseline 2 for $\VUE=100$ VUEs.
However, the average latency of Baseline 2 outperforms the proposed for $\VUE=$ 20 and 60 while being slightly higher when $\VUE=100$.
The main advantage of the proposed method is the reduction in the extreme cases where VUEs exceed the queue threshold.
The proposed method reduces the fraction of VUEs exceeding queue threshold by 60\%, 8\%, and 9\% compared to Baseline 2 without additional transmit power on average for $\VUE=$ 20, 60, and 100, respectively. 
Compared to the fixed power and Baseline 1 models, the reductions in the fraction of VUEs exceeding queue threshold are as high as 97-99\%.

\section{Conclusions}\label{sec:conclusion}
In this paper, we have formulated the problem of joint power control and resource allocation for V2V communication network as a network-wide power minimization problem subject to ultra reliability and low latency constraints.
The constraints on URLLC are characterized using extreme value theory and modeled as the tail distribution of the network-wide queue lengths over a predefined threshold.
Leveraging concepts of federated learning, a distributed learning mechanism is proposed where VUEs estimate the tail distribution locally with the assistance of a RSU.
Here, FL enables VUEs to learn the tail distribution of the network-wide queues locally without sharing the actual queue length samples reducing unnecessary overheads.
Combining both EVT and FL approaches, we have proposed a Lyapunov-based distributed transmit power and resource allocation procedure for VUEs.
Using simulations, we have shown that the proposed method learns the statistics of the network-wide queues with high accuracy.
Furthermore, the proposed method shows considerable gains in reducing extreme events where the queue lengths grow beyond a predefined threshold compared to systems that account for reliability by imposing probabilistic constraints on the average queue lengths.

\appendices
\section{Proof of Proposition \ref{prop:derivative_coefficient}}\label{appndx:gevd_gradient}

Let $\gpdExponent{\sample} = (1 + \gpdShape \sample / \gpdScale)^{-1/\gpdShape}$. 
Since $\gpdExponent{\sample} \to e^{- \sample / \gpdScale}$ as $\gpdShape \to 0$, the distribution can be rewritten as $\gpdML(\sample) = \frac{1}{\gpdScale}\gpdExponent{\sample}^{\gpdShape + 1}$.
Using the above notation, it can be noted that, 
\begin{equation}
	\loglikelihood(\sampleSet)
	= \frac{1}{\setSize{\sampleSet}}\sum_{\sample\in\sampleSet} \Big(  \ln \gpdScale - (\gpdShape+1) \ln \gpdExponent{\sample} \Big) 
	= \frac{1}{\setSize{\sampleSet}}\sum_{\sample\in\sampleSet} \loglikelihood(\sample).
\end{equation}
Hence, $\grad{\gpdCombined}{\loglikelihood(\sampleSet)} = \frac{1}{\setSize{\sampleSet}}\sum_{\sample\in\sampleSet} \grad{\gpdCombined}{\loglikelihood(\sample)}$ is held.

First, the gradient of $\gpdExponent{\sample}$ is found by,
\begin{equation}\label{eqn:gradients_exponent_term}
\grad{\gpdCombined}{\gpdExponent{\sample}}
= \begin{bmatrix}
\daba{\gpdExponent{\sample}}{\gpdScale} \\
\daba{\gpdExponent{\sample}}{\gpdShape}
\end{bmatrix}
= \begin{bmatrix}
\frac{\sample}{\gpdScale} \gpdExponent{\sample}^{\gpdShape+1}  \\
\frac{\gpdExponent{\sample}}{\gpdShape^2} \big( \gpdExponent{\sample}^{\gpdShape} - \gpdShape \ln \gpdExponent{\sample} -1 \big)
\end{bmatrix}.
\end{equation}
Thus, the gradient of $\loglikelihood(\sample)$ can be calculated as follows:
\begin{align*}
\grad{\gpdCombined}{\loglikelihood(\sample)}
&= \begin{bmatrix}
\daba{\loglikelihood(\sample)}{\gpdScale} \\
\daba{\loglikelihood(\sample)}{\gpdShape}
\end{bmatrix}
= \begin{bmatrix}
\frac{1}{\gpdScale} - \frac{1 + \gpdShape}{\gpdExponent{\sample}} \daba{\gpdExponent{\sample}}{\gpdScale}  \\
- \frac{1 + \gpdShape}{\gpdExponent{\sample}} \daba{\gpdExponent{\sample}}{\gpdShape} - \ln \gpdExponent{\sample}
\end{bmatrix} \\
&= \begin{bmatrix}
\frac{1}{\gpdScale} \big( \frac{1 + 1/\gpdShape}{ 1 + \gpdShape \sample / \gpdScale } - \frac{1}{\gpdShape} \big)\\
\frac{(1 + 1/\gpdShape) ( 2 + \gpdShape \sample / \gpdScale )}{ 1 + \gpdShape \sample / \gpdScale } - \frac{ \ln (1 + \gpdShape \sample / \gpdScale) }{\gpdShape^2} 
\end{bmatrix}.
\end{align*}

\bibliographystyle{IEEEtran}
\bibliography{IEEEabrv,mybib_v2x}

\end{document}